\documentstyle[12pt]{article}

\newcommand{\beq}{\begin{equation}}
\newcommand{\eeq}{\end{equation}}
\newcommand{\bea}{\begin{eqnarray}}
\newcommand{\eea}{\end{eqnarray}}

\newcommand{\gsim}{\lower.7ex\hbox{$
\;\stackrel{\textstyle>}{\sim}\;$}}
\newcommand{\lsim}{\lower.7ex\hbox{$
\;\stackrel{\textstyle<}{\sim}\;$}}

\def\ot{{\bf T}}
\def\cp{{\bf CP}}
\def\cpt{{\bf CPT}}

\newcommand{\GeV}{\,\mbox{GeV}}
\newcommand{\MeV}{\,\mbox{MeV}}
\newcommand{\matel}[3]{\langle #1|#2|#3\rangle}

\newcommand{\eod}{\end{document}}

\begin{document}
\thispagestyle{empty}
\vspace*{-20mm}

\begin{flushright}
UND-HEP-06-BIG\hspace*{.08em}07\\
hep-ph/0608225\\


\end{flushright}
\vspace*{7mm}

\begin{center}
{\LARGE{\bf
The Physics of Beauty (\& Charm [\& $\tau$]) at  
\vspace*{2mm}
the LHC and in the Era of the LHC }}
\footnote{Invited Plenary Lecture given at {\em Physics at LHC}, Cracow, Poland, July 2006}
\vspace*{14mm}

{\large{\bf I.I.~Bigi}}\\
\vspace{4mm}

 {\sl Department of Physics, University of Notre Dame du Lac}
\vspace*{-.8mm}\\
{\sl Notre Dame, IN 46556, USA}\vspace*{1.5mm}\\
email: ibigi@nd.edu

\vspace*{5mm}

{\bf Abstract}\vspace*{-.9mm}\\
\end{center}

\noindent
The recent successes of the SM do not weaken the arguments in favour of New Physics 
residing at the TeV scale. Finding and identifying it represents the prime challenge for a generation of high energy physicists. To differentiate between different scenarios of New Physics we need to analyze their impact on flavour dynamics. 
A continuing comprehensive program of heavy flavour studies instrumentalizing the high sensitivity 
of \cp~analyses  is intrinsically connected to LHC's core mission.  In $B$ decays we 
can typically expect no more than moderate deviations from SM predictions. $B_s$ transitions provide an {\em autonomous} access to New Physics 
not prejudiced by $\Delta M(B_s)|_{exp}\simeq \Delta M(B_s)|_{SM}$. 
Dedicated studies of charm and 
$\tau$ decays offer unique opportunities to observe New Physics. One challenge is whether 
LHCb will be able to exploit 
LHC's huge charm production rate to probe for \cp~asymmetries. Likewise, to which degree 
ATLAS/CMS can contribute to $B$ physics and to searches for $\tau \to 3l$.  Yet to saturate the discovery potential for New Physics in beauty, charm and $\tau$ decays we will need a comprehensive 
high quality data base that only a Super-Flavour Factory can provide.  

\vspace*{10mm}
\section{Introduction}
Around the turn of the Millenium we have experienced a `quantum jump' in knowledge, though not in 
understanding:  
\begin{itemize}
\item 
The Standard Model (SM) Paradigm of Large \cp~Violation in $B$ decays has been validated. 
\item 
$\nu$ oscillations have been established experimentally (and the solar model validated as well in the 
process). 
\item 
Evidence for `Dark Energy' has emerged -- a concept concisely characterized by the 
quote:"Who ordered that?"
\end{itemize}
Even the first item -- a great, unqualified and novel success of the SM -- does {\em not} invalidate 
the arguments in favour of the SM being incomplete already around the TeV scale. 

This is not a surprising statement for the audience at this conference. For the central justification 
for the LHC is to reveal the dynamics driving the electroweak phase transition. Our 
foremost goal has to be to make the LHC succeed greatly -- even beyond our expectations -- 
and in the process prove Samuel Beckett wrong who said: "Ever tried? Ever failed? No matter. 
Try again. Fail again. Fail better." The second and third item above tell us we will not fail forever; 
furthermore a `New \cp~Paradigm' is needed to implement baryogensis. 
I am actually confident that we will `succeed' soon. 

My central message can be summarized as follows: We must study the impact of that anticipated 
New Physics on flavour dynamics. The LHCb program is thus {\em intrinsically connected} to the 
{\em core mission} of the LHC. The required comprehensive flavour studies have to include the charm and 
$\tau$ lepton sector. The goal here is not primarily to enlighten us about the flavour mystery -- 
although that could come about -- and not to unveil the New \cp~Paradigm needed for baryogenesis -- 
though it can quite conceivably happen --  but to 
{\em instrumentalize} the high sensitivity inherent in 
\cp~studies to interpret the footprints of New Physics to be revealed in high $p_{\perp}$ studies.  Dedicated and comprehensive flavour studies are a necessity -- not a luxury -- in the 
Era of the LHC, and I view a Super-Flavour Factory a most desirable component of it. 

After an update on the SM's paradigm of large \cp~violation 
in $B$ decays in Sect.\ref{PARA}  I discuss the lifetimes of beauty hadrons in  
Sect.\ref{BLIFE} and sketch future $B$ studies in Sect.\ref{FUTURE};  $D$ and $\tau$ decays are addressed in Sect.\ref{DARK} before concluding with a plea for a Super-Flavour Factory; some  
more technical comments on Heavy Quark Theory and Dalitz plots analyses are shifted to Appendices.

\section{The SM's Paradigm of Large \cp~Violation in $B$ Decays -- a Triple Triumph}
\label{PARA}

Three central consequences of CKM theory had been {\em pre}dicted: 
\begin{enumerate}
\item 
Some $B$ decay modes like $B_d \to \psi K_S$, $B_d \to \pi^+\pi^-$ and 
$B \to K^+\pi^-$ have to exhibit truly large \cp~asymmetries -- there is no 
"plausible deniability". 
It is very nontrivial to infer from a \cp~asymmetry in the $K^0 - \bar K^0$ system measured to be 
on the few$\times 10^{-3}$ level that $B$ decays should exhibit \cp~violation hundred times  
larger, i.e. close to the largest values mathematically possible \cite{BS80}.  
\item 
Large {\em direct} \cp~violation has to occur as well \cite{SONI}. 
\item 
The magnitudes of \cp~{\em in}sensitive observables -- $|V(ub)/V(cb)|$ and $\Delta M(B_d)/\Delta M(B_s)$ --  control the existence and strength of \cp~violation, as expressed through 
$\epsilon_K$ and sin2$\phi_1$.  
\end{enumerate} 
All these predictions have now been validated experimentally \cite{BART}. Since the summer of 2001 we 
can say: (i) The CKM paradigm has become a {\em tested} theory. (ii) \cp~violation has been 
{\em de}mystified: if the dynamics is sufficiently complex to support \cp~violation, there is no 
a priori reason, 
why the latter should be small; i.e. weak complex phases can be large. (iii) The de-mystification 
will be completed -- a good thing in my view -- if \cp~violation is found anywhere in leptodynamics. 

While there are certain regions in kaon  dynamics with large \cp~asymmetries -- the 
interference region in $K_{neut} \to \pi \pi$ or the \ot~odd correlation between the 
$\pi^+\pi^-$ and $e^+e^-$ planes in $K_L \to \pi^+\pi^-e^+e^-$ -- the statement that 
"\cp~violation in $B$ decays is much larger than in $K$ decays" is an empirically verified 
fact: while the $K_L$ (and $K_S$) act like \cp~eigenstates to a very good approximation, 
this not at all true for the mass eigenstates of the $B_d -\bar B_d$ system. 

To summmarize the 2006 status more quantitatively \cite{BART,ICHEP06}: 
\begin{itemize}
\item 
From $B_d \to \psi K_S$ one obtains  
\beq 
{\rm sin}2\phi_1|_{WA} = 0.674 \pm 0.026   \; \; vs. \; \; {\rm sin}2\phi_1|_{CKM} = 0.725 \pm 0.065
\eeq
The `battle for supremacy' has been decided: we search no longer for {\em alternatives} 
to CKM theory, but for {\em corrections}. At the same time baryogenesis has to be driven by 
dynamics other than CKM; thus we can be confident that CKM forces do not represent a monopoly. 
\item 
Direct \cp~violation has been established in $B_d \to K^+\pi^-$ by both BABAR and BELLE.   

While the BABAR and BELLE data sets on  $B_d \to \pi^+\pi^-$ do not form a perfect union (yet), 
the BELLE analysis shows large \cp~violation of the indirect as well as direct variety. A 
time-dependent \cp~asymmetry in $B_d \to 2\pi $ can be expressed as a sum of  
sin and cos$\Delta M(B_d)$ terms with coefficients $S$ and $C$, respectively. 
With{\em out direct} \cp~violation one obviously has $C = 0$; yet in addition also 
$S = - {\rm sin} 2\phi_1 \simeq - 0.7$ has to hold,  where the minus sign is due to the 
$2\pi$ and $\psi K_S$ final states having opposite \cp~parity. I.e., once \cp~violation in 
$B_d \to \psi K_S$ has been established, one infers the existence of {\em direct} \cp~violation 
from $(S,C) \neq (-{\rm sin}2\phi_1,0)$ rather than from $(S,C) \neq (0,0)$.

\item 
Recently both the D0 and CDF collaborations  
reported a signal for $B_s - \bar B_s$ oscillations  \cite{D0BSOSC,CDFBSOSC}: 
\beq 
\Delta M(B_s) = 
\left\{   
\begin{array}{ll}
(19 \pm 2 ) \, {\rm ps}^{-1} & {\rm D0} \\
(17.3 ^{+0.42}_{-0.21} \pm 0.07)\, {\rm ps}^{-1} & {\rm CDF} \\
(18.3 ^{+6.5}_{-1.5}) \, {\rm ps}^{-1} & {\rm CKM\; fit} \\
\end{array} 
\right. 
\label{DIRECTCPDATA} 
\eeq 
While the strength of the signal has not yet achieved 5 $\sigma$ significance, it looks most 
intriguing. {\em If} true, it represents another impressive triumph of CKM theory: 
the \cp~{\em in}sensitive observables $|V(ub)/V(cb)|$ and $\Delta M(B_d)/\Delta M(B_s)$ -- 
i.e. observables that do {\em not} require \cp~violation for acquiring a non-zero value -- imply 
(a) a non-flat CKM triangle and thus \cp~violation, see the left of Fig.~\ref{fig1},   
that (b) is fully consistent with the observed \cp~sensitive observables $\epsilon_K$ 
and sin$2\phi_1$, see the right of Fig.~\ref{fig1}.
\begin{figure}[t]
\vspace{5.0cm}
\includegraphics{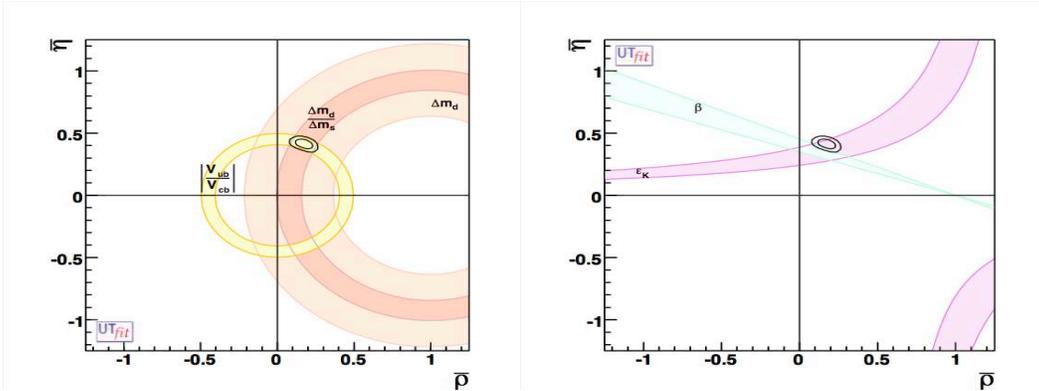}
\caption{Unitarity triangle from $|V(ub)/V(cb)|$ \& $\Delta M(B_d)/\Delta M(B_s)$ on the left and  
       compared to constraints from 
$\epsilon_K$ \& sin2$\phi_1/\beta$ on the right 
(courtesy of M. Pierini)
    \label{fig1} }
\end{figure}

\end{itemize}
These successes of the SM tell us that we can{\em not count} on {\em numerically} massive manifestations 
of new dynamics in beauty transitions. Accordingly we must strive to achieve as high an 
accuracy level in our theoretical description as possible.  The goal of high accuracy is not utopian -- it can be achieved by combining a {\em robust} theoretical 
framework with {\em comprehensive} data as illustrated in the Appendices.

\section{Lifetimes of Beauty Hadrons}
\label{BLIFE}

Based on the Heavy Quark Expansion (HQE), which is sketched in Append.\ref{HQTH}, 
the lifetime ratios of beauty hadrons have been 
predicted in the old-fashioned sense, i.e. {\em before} meaningful measurements had been 
undertaken. In Table \ref{tab:TABLEBEAUTY} I list the original {\em pre}dictions 
together with later updates and the data. 
\begin{table}
\begin{center}
\small{\begin{tabular}{lll}
\hline 
 & $1/m_b$ expect.  & data   \\
 \hline 
\hline
$\frac{\tau (B^+)}{\tau (B_d)}$ &  
$\sim 1+ 0.05\left( \frac{f_B}{200\; \MeV} \right)^2$    '92 \cite{MIRAGE}  
             & $1.076 \pm 0.008$  \cite{WINTER06} \\    
 & $1.06 \pm 0.02$    \cite{LENZ}  & \\ 
 \hline 
$\frac{\overline{\tau} (B_s)}{\tau (B_d)}$ & $1 \pm {\cal O}(0.01)$   '94 \cite{DSSL} 
& $0.958 \pm 0.039$ \cite{WINTER06} \\
\hline 
 $\frac{\tau (\Lambda _b^-)}{\tau (B_d)}$ & $\geq 0.9 $    '93 \cite{STONEBOOK}
     & $0.806 \pm 0.047$   WA '05 \cite{WINTER06} \\   
 & $\simeq 0.94$ \& $\geq 0.88$  '96 \cite{BOOST,FAZIO} &  $1.037\pm 0.058$   CDF '06 
 \cite{CDFNOTE} \\
 & & $0.870 \pm 0.102 \pm 0.041$  D0 '06\cite{D045}\\
 \hline
$\tau (B_c)$ & $\sim (0.3 - 0.7)$ psec   '94ff  \cite{MICHEL}   & 
$0.45 \pm 0.12$ psec  
\cite{WINTER06}\\
 &  & \\
 \hline   
$\frac{\Delta \Gamma (B_s)}{\overline{\Gamma}(B_s)}$ & 
$22\% \cdot  \left(\frac{f(B_s)}{220\, \MeV}\right) ^2$  '87 \cite{AZIMOV} & 
$0.65 \pm 0.3$  CDF \\
 & $12 \pm 5\% $   '04 \cite{LENZ} &  $0.26 \pm 0.14$  \cite{D039} \\
\hline
\end{tabular}}
\caption{Weak lifetime ratios of beauty hadrons}
\label{tab:TABLEBEAUTY}
\end{center}
\end{table}
Several comments are in order: (i) The prediction on $\tau (B^+)/\tau(B_d)$ is in 
pleasing agreement with rather accurate data. 
(ii) The largest deviation from uniform lifetimes occurs 
for $B_c$ mesons: their lifetime is that of charm hadrons as expected in a naive 
additive quark model picture and predicted by the HQE, where the absence of a $1/m_Q$ 
correction is essential. (iii) The long saga on the $\Lambda_b$ lifetime appears to have taken 
a surprising turn. The authors of the original prediction and its refinement 
`stuck to their guns' about $\tau (B_d)$ exceeding $\tau (\Lambda_b)$ by not significantly 
more than 10 \%, when for several years the data seemed to clearly indicate otherwise: as 
late as 2005 the world average still read $\tau (\Lambda_b)/\tau (B_d)= 0.806 \pm 0.047$.  
During that time other theorists gave different predictions \cite{PETROV}. 
I am eagerly awaiting future data, in particular also on $\tau (\Xi_b^{0,-})$ 
\cite{VOLOSHINXIB}. (iv) The prediction that the average lifetime of the $B_s$ mass eigenstates 
should differ from $\tau (B_d)$ by merely a percent or two is a carefully analyzed, yet not an 
ironclad one. Previous data indicated a somewhat lower value, which would also have been 
more consistent with the first results on $\Delta \Gamma (B_s)$, see below. Yet it appears 
data are moving closer to the original theoretical prediction.  (v) The measured values for 
$\Gamma (B_s \to D_s^{(*)}\bar D_s^{(*)})$ can give a reasonable ballpark estimate for 
$\Delta \Gamma (B_s)$; yet a real prediction is best obtained by evaluating 
the `quark-box diagram' \cite{AZIMOV,LENZ}. Two qualifying remarks are important though: 
($\alpha$) A ratio of 0.25 is almost a `unitarity' bound on 
$\Delta \Gamma (B_s)/\Gamma (B_s)$, unless $\bar \Gamma (B_s)$ is significantly larger 
than $\Gamma (B_d)$ contrary to expectations, see above. 
($\beta$) While the quark box diagrams used for evaluating 
$\Delta M$ as well as $\Delta \Gamma$ look very similar, the dynamical situation is quite 
different in the two cases. $\Delta M$ is controlled by {\em off}-shell transitions and thus involves a 
considerable amount of averaging over hadronic channels making duality a good approximation 
\cite{VADE}. 
On the other hand $\Delta \Gamma$ is given by {\em on}-shell transitions with less averaging, 
which could enhance the limitations to duality considerably. Furthermore $\Delta \Gamma$ as 
obtained from the quark box diagram is considerably reduced by GIM cancellations; however those 
could be modified very significantly for the on-shell modes due to the proximity of the 
$D_s^{(*)}\bar D_s^{(*)}$ thresholds. I am {\em not} suggesting that employing the quark box diagram for evaluating 
$\Delta \Gamma (B)$ is unreasonable. I am concerned about the following:  while in the ratio 
$\Delta M(B_s)/\Delta \Gamma (B_s)$ some uncertainties like bag factors and decay constants 
cancel, $\Delta \Gamma (B_s)$ might suffer from further theoretical uncertainties, which could significantly bias the prediction for it as well as for $\Delta M(B_s)/\Delta \Gamma (B_s)$.

The HQE has been successful even on a quantitative level in predicting the lifetimes 
of beauty hadrons, which after all are dominated by nonleptonic transitions. The basic feature 
that nonperturbative corrections arise first in order $1/m_b^2$ holds for both semileptonic and 
nonleptonic widths \cite{BUV}, yet in the latter there are more perturbative QCD corrections, and limitations to quark-hadron duality are likely to be somewhat larger on average (and possibly significantly 
larger in some cases).

\section{On Future Lessons in $B$ Decays}
\label{FUTURE}

Since we can{\em not 
count on quantitatively massive} manifestations of New Physics in $B$ decays, we 
must combine high accuracy with high sensitivity in heavy flavour studies. Some tools  
for attaining  such a goal   
are briefly addressed in Append.~\ref{HQTH} - \ref{DALITZ}.

\subsection{Rare $B$ Decays}

$\Gamma (B\to l \nu X_c)$: the {\em inclusive} semileptonic width has been calculated with about 3\% theoretical uncertainty for 
$l = e,\mu$ \cite{BENSON1}. The only evaluation for $l = \tau$ has been given twelve years ago 
\cite{FALKNIR}, and the only 
measurement is equally old. Now we have the tools to 
compute $\Gamma (B \to \tau \nu X_c)/\Gamma (B \to e\nu X_c)$ much more precisely. Measuring it with commensurate precision allows a search for New Physics, 
in particular in the form of an extended Higgs sector, since charged Higgs exchange would affect 
$B\to \tau \nu X_c$ most significantly. 

\noindent A variant of such a probe is to compare the {\em exclusive} 
rates for   
$B \to \tau \nu D$ vs. $B \to e \nu D$ \cite{TANAKA}, since the former unlike the latter could be affected 
by a charged Higgs as heavy as several hundred GeV. There is one complication, though: contrary to claims in the literature hadronization effects do {\em not} drop out from the ratio. Yet this problem 
can be overcome through Uraltsev's `BPS' approximation \cite{BPS}, as explained in 
Append.~\ref{BPSAP}.

$B \to l\bar l X$: While the {\em inclusive} transitions can be measured only at a $B$ factory, 
the {\em exclusive} channels $B \to l^+l^- K/K^*$ -- rates, lepton spectra, 
\cp~and forward-backward asymmetries -- can be studied at the LHC, in particular by 
LHCb with its superb particle id. 

\noindent Urging to measure $\Gamma (B \to \nu \bar \nu X_s)$, even exclusively, is {\em not}  
the result of the frivolous nature of theorists \cite{BUCHETAL}. For the dynamical information to be gained 
here is in general quite independent from that in $B \to l^+l^-X_s$. Alas -- it is in the domain of a 
Super-$B$ factory.

\subsection{Flavour-Changing Neutral Currents in $B_s$ Decays -- an Independent Chapter 
in Nature's Book on Fundamental Dynamics}

The $B$ factories have been much more successful than anticipated in the quality of their measurements. Among many other achievements they have determined the three angles of the 
CKM triangle with higher accuracy than expected from $B_{d,u}$ decays. We have to focus now on finding and subsequently identifying non-CKM corrections. 

Originally it was thought that $B_s$ decays are needed in an essential way to construct the 
CKM triangle, namely to extract the angle $\phi_3$ and the side $|V(td)/V(ts)|$. For the 
latter this is true, as mentioned above. Yet the angle $\phi_3$ is being determined with 
good accuracy in $B^{\pm} \to D^{\rm neut}K^{\pm}$. I view the statement that we will extract 
$\phi_3$ from $B_s$ as somewhat missing the point. Our primary goal is to search for New Physics. While within the SM similar quark box diagrams affect rare 
transitions and oscillations for $B_d$ and $B_s$ decays, we have to `think outside the box' -- 
pun intended. For a priori there is no reason why New Physics should affect 
$B_s$ and $B_d$ transitions with similar weights as within the SM. $B_s$ channels should therefore 
be analyzed in an {\em autonomous} way. 

$\Delta M(B_s)$ and $\Delta \Gamma (B_s)$ have been addressed already. 
I will give four examples where New Physics can still impact on $B_s$ decays in a numerically massive way: 
\begin{itemize}
\item 
The rate for $B_s \to \mu^+\mu^-$ with 
BR$(B_s \to \mu^+\mu^-)|_{SM} \sim 3\cdot 10^{-9}$ can be greatly enhanced 
in some SUSY scenarios by $({\rm tg}\beta )^6$ \cite{KOLDA}, which could produce a rate right at the 
experimental upper bound of $10^{-7}$. 
\item 
The time-dependent \cp~asymmetries in $B_s \to \psi \phi/\psi \eta ^{(\prime)}$ are reliably predicted 
to be small in the SM \cite{BS80}, namely below 4 \%. For on the leading Cabibbo level only quarks of the 
second and third family contribute, and by themselves they cannot induce \cp~violation. Yet New Physics could produce a \cp~asymmetry as large as several $\times 10$ \% -- even with 
the observed value of $\Delta M(B_s)$ close to the SM prediction. 
\item 
With oscillations leading to `wrong-sign' leptons -- $\bar B_s \to l^+ \nu X$ and $B_s \to l^- \nu X$ -- 
one can probe for a \cp~asymmetry there. Within the SM it has to be tiny $\sim {\cal O}(10^{-4})$, 
since it is suppressed by $\Delta \Gamma /\Delta M$ and by the leading contributions again coming 
from quarks of only the second and third family. Yet the second suppression factor could be 
vitiated by New Physics leading to a semileptonic \cp~asymmetry two orders of magnitude larger. 
\item 
The mode $B_s \to \phi \phi$ is the analogue of $B_d \to \phi K_S$: within the SM its \cp~asymmetry   
has to basically coincide with that of $B_s \to \psi \phi$, i.e. be very small; yet since it is driven by 
a one-loop process, i.e. with a suppressed SM amplitude, it is quite susceptible to New Physics. 
Ultimately it offers one intrinsic advantage over $B_d \to \phi K_S$: once one has differentiated the 
contributions from $l=0,1,2$ partial waves, one can analyze in which partial wave a possible direct \cp~asymmetry arises. LHCb will be particularly well suited for this task.   
\end{itemize}

\subsection{On the Capabilities of Hadronic Collider Experiments}

While CMS, ATLAS and LHCb should be able to search for $B_s \to \mu^+\mu^-$, it is not clear, 
if even LHCb can probe for $B_s \to \tau^+\tau^-$, despite the latter's branching ratio being larger by two orders of magnitude. 

The relatively low value reported by CDF/D0 for $\Delta M(B_s)$ should allow also ATLAS and CMS to track $B_s$ oscillations. Whether this will yield 
enough sensitivity to hunt (time-dependent) \cp~asymmetries in $B_s \to \psi \phi/\eta$ or in 
$B_s \to D_s K$ with the latter [former] probing for New Physics in $\Delta B=1\&2$ 
[$\Delta B =2$] dynamics, will depend on the quality of the flavour tagging and particle id; likewise 
for $\bar B_s \to l^+X$ vs. $B_s \to l^-X$. To be able to study rates, lepton spectra and asymmetries 
in $B \to l^+l^- K/K^*/\pi /\rho$ and $B_s \to l^+l^-\eta/\phi, K^*$ will require efficient triggers, good 
flavour tagging and particle id. I expect LHCb to be well up to the task.

\section{The Dark Horses -- Charm Quarks and $\tau$ Leptons}
\label{DARK}

$B$ decays (and similarly for kaons) with their large CKM suppression are a most natural place to search for New Physics. Yet we have to search in 
unconventional places as well.

\subsection{On the Future Promise of Charm}

Accurate measurements of leptonic as well as semileptonic charm decays will teach us novel 
lessons about nonperturbative QCD, calibrate and hopefully validate our theoretical tools that 
then can be employed with more confidence in $B$ studies. This is the foundation of the 
CLEO-c program. Yet there is much more beyond this `guaranteed profit': New Physics could induce 
flavour changing neutral currents that are considerably less suppressed for up- than for down-type 
quarks. Only 
charm allows the full range of probes for New Physics in general and flavour-changing 
neutral currents in particular: 
(i) Since top quarks do not hadronize \cite{RAPALLO}, there can be no $T^0- \bar T^0$ oscillations. More generally, hadronization, while hard to bring under theoretical control, enhances the 
observability of \cp~violation \cite{RIO}. 
(ii)  
As far as $u$ quarks are concerned, $\pi^0$, $\eta$ and $\eta ^{\prime}$ decays electromagnetically, not weakly. They are their own antiparticles and thus cannot oscillate. \cp~asymmetries are mostly 
ruled out by \cpt~invariance. 

My basic contention: {\em Charm transitions provide a unique portal 
for finding the intervention of New Physics in flavour dynamics with the experimental situation being a priori quite 
favourable (apart from the absence of Cabibbo suppression). Yet even that handicap can be overcome 
by statistics. }

I am quite skeptical that the observation of $D^0 - \bar D^0$ oscillations by themselves can establish 
the intervention of New Physics, since the SM predictions for $x_D= \Delta M_D/\Gamma_D$ and 
$y_D = \Delta \Gamma _D/2\Gamma_D$ yield values $\sim {\cal O}(10^{-3})$ 
\cite{BUDDBAR} -- and might allow even 
$10^{-2}$ \cite{FALKDDBAR} -- when the data read $x_D \leq 0.03$ and $y_D = 0.01 \pm 0.005$. Nevertheless one should make every effort to observe it, mainly because it can provide independent validation 
for a signal of \cp~violation involving $D^0 - \bar D^0$ oscillations. Such an effect would represent 
conclusive proof for the intervention of New Physics. 

Since baryogenesis implies the existence of New Physics in \cp~violating dynamics, we better  undertake dedicated searches for \cp~asymmetries in 
charm decays, where the `background' from known physics is between absent and small 
\cite{CICERONE,CHARM06}. 
Most experimental facts help a search for \cp~violation due to New Physics, be it of the direct or 
indirect variety, be it in partial widths or final state distributions. I will list just two examples: 
\begin{itemize}
\item 
One can search for a time-dependent difference in the rates for the doubly Cabibbo 
suppressed modes $D^0 \to K^+\pi^-$ vs. $\bar D^0 \to K^- \pi^+$ \cite{BERKICHEP}. With LHCb expecting to record about 
$5 \cdot 10^7$ {\em tagged} $D^* \to D + \pi  \to K^+K^- +\pi$ events in a nominal year 
of $10^7$ s \cite{TAT}, one will achieve very high sensitivity for New Physics.
\item 
In $\stackrel{(-)}D \to K \bar K \pi^+\pi^-$ one can measure the angle $\phi$ between the 
$K \bar K$ and $\pi^+\pi^-$ planes and probe for a difference in the $\phi$ distribution for 
$D$ and $\bar D$ decays. Since one can measure \ot~odd and even moments separately 
for $D$ and $\bar D$, one should be able to control systematics in the detection efficiencies of 
particles and antiparticles \cite{CHARM06}.

\end{itemize}

\subsection{$\tau$ Decays -- an almost Unique Opportunity}

Lepton flavour violating (LFV) modes like $\tau \to l \gamma/3l$ with $l=e,\mu$ require 
New Physics to occur. While searching for $\tau \to l \gamma$ appears beyond the capabilities of the LHC, $\tau \to 3 l$ with its present upper bound BR$(\tau \to 3 l) \leq {\rm few}\times 10^{-7} $ 
does not. For semileptonic $B$ transitions produce about 
${\cal O}(10^{12})$ $\tau$ leptons per year. While the width for $\tau \to 3 l$ tends to be smaller than for $\tau \to l \gamma$ in most New Physics models, there 
are exceptions; more importantly the former is typically within an order of magnitude of the 
latter. If for illustrative purposes one makes two ad-hoc assumptions, namely that 
(a) New Physics makes up half of the observed $B_d \to \phi K_S$ amplitude and (b) the corresponding 
lepton coupling is equal in size, one arrives at 
BR$(\tau \to 3 \mu) \sim {\cal O}(10^{-8})$ after this crude exercise. 

An even more ambitious task is to probe for \cp~violation in $\tau$ decays. As already mentioned 
a new source of \cp~violation is needed to implement baryogenesis. Furthermore {\em lepto}genesis 
might be the primary process; in that case it is essential to identify \cp~violation in leptodynamics. 
I see a realistic chance for success in three areas only: neutrino oscillations, the electric dipole moment 
of electrons -- and $\tau$ decays, in particular in the channels $\tau \to \nu K\pi$. For 
while those are Cabibbo suppressed in the SM, they should be particularly sensitive to exchanges of 
charged Higgs bosons. A \cp~asymmetry can arise not merely in the partial widths -- and known dynamics has to induce a $0.0032$ asymmetry in $\tau \to \nu K_S\pi$ \cite{BSTAU} -- , but also in the 
final state distributions \cite{KUEHN}; the $\tau$ spin can be used as a powerful observable in 
$e^+e^- \to \tau^+\tau^-$ by employing the spin alignment of the $\tau$ pair or -- better still -- having 
the electron beam polarized. None of this can be achieved at the LHC. Since an optimistic, yet 
not  unrealistic range is given by the $10^{-3}$ level 
\cite{SWEDEN}, this is a noble task for a Super-Flavour Factory. 

For proper perspective one should note that the rates for LFV modes are quadratic in New Physics 
amplitudes; \cp~asymmetries in $\tau$ decays, on the other hand, have to be only linear, since 
the SM provides the other amplitude. Searching for a LFV rate on the $10^{-8}$ level is thus 
of comparable sensitivity to New Physics as a \cp~asymmetry of order $10^{-3}$ in a Cabibbo suppressed mode.

\section{Summary \& Outlook -- on the Need for a Super-Flavour Factory}
\label{OUT}

There have been many good news for the SM over the last five years. In particular its 
paradigm of large \cp~asymmetries -- both indirect and direct ones -- in $B$ decays has been validated. Through CKM dynamics the SM provides at least the lion's share of the 
\cp~asymmetries observed in $K_L$ and $B$ decays. The SM appears to have scored another 
impressive success of a new quality with an observable given purely by quantum corrections: 
the value seen for $\Delta M(B_s)$ is quite consistent with the prediction, and together with 
another \cp~{\em in}sensitive quantity -- $|V(ub)/V(cb)|$ -- it constrains the \cp~observables 
$|\epsilon _K|$ and sin$2\phi_1$ very close to their measured values. While giving theorists 
the unwelcome feeling of "deja vue all over again", it represents good news for dedicated 
experimentalists.  For even with $\Delta M(B_s)|_{exp} \simeq \Delta M(B_s)|_{SM}$ the 
time-dependent \cp~asymmetry in $B_s \to \psi \phi/\eta$ can still exceed the small value predicted by the SM 
by an order of magnitude. In addition the `moderate' value of $\Delta M(B_s)$ should allow 
ATLAS and CMS to participate in the hunt for New Physics through resolving the oscillations 
in $B_s \to \psi \phi$ and $B_s \to l^+l^-\phi$. While in these processes and a few others like 
$B_s \to \mu^+\mu^-$ and $B_s \to l^-X$ vs. $\bar B_s \to l^+X$ the deviations from SM expectations can be large, I expect New Physics to induce typically smallish effects only. Thus high premium has to be placed 
on accuracy on the experimental as well as theoretical side, the latter concerning making predictions and interpreting the data. Heavy quark theory and its $1/m_Q$ expansions implemented through the OPE, augmented by (hopefully) validated lattice QCD and calibrated by a large body of `flanking' measurements should allow us to attain this ambitious goal. These elements are sketched in the 
Appendix below. 

I can hardly over-emphasize that $B_s$ transitions represent an independent chapter in Nature's 
Book on Fundamental Dynamics and therefore fully deserve a comprehensive and detailed program 
of research. 

A large discovery potential for New Physics exists also in $\tau$ decays -- LFV and \cp~violation -- 
and in weak charm decays mainly through \cp~studies. We know the SM can{\em not} implement 
baryogenesis. Charm is the only up-type quark that allows a full probe of New Physics 
through flavour changing neutral currents, and only recently have we entered a  domain with a realistic chance to see something novel. 

There arise two questions to the LHC community: (i) Can LHCb 
take up the charm challenge, i.e. trigger with sufficient efficiency on charm to exploit the statistical muscle of the LHC for high quality charm studies? (ii) Can ATLAC \& CMS go after 
$\tau \to 3 l$?     

Let me conclude with some glimpses of the `Big Picture'. I am confident that there resides indeed 
New Physics around the TeV scale ({\em cpNP}), and LHC will find its footprints. Identifying its features has to be our central goal. This {\em cpNP} can 
affect flavour dynamics {\em significantly}, though not necessarily massively. Analyses of heavy flavour 
decays -- in the quark as well as lepton sector -- are likely to reveal some of these salient features and thus provide probes of the {\em cpNP} complementary to high $p_{\perp}$ observations. I view a continuing dedicated program of heavy flavour studies essential -- {\em not} a luxury -- where the 
high sensitivity of \cp~studies in particular is mainly {\em instrumentalized} to probe the {\em cpNP}. 

To saturate the discovery potential in $B$ decays we need numerically reliable and precise tools. In the last fifteen years we have made great strides in that respect by developing various aspects of heavy 
quark theory and will continue to do so. One lesson we can take from there is that the availability of 
precise data and the challenges provided by them drive (at least some) theorists to strive for higher accuracy.  

In charm and $\tau$ decays on the other hand one can hope for numerically massive deviations from 
SM predictions, since the latter are a bit on the dull side; yet one has to push the experimental sensitivity as high as possible. 

LHC's high $p_{\perp}$ program represents largely hypothesis-probing research; $B$ studies have 
significant aspects also of  hypothesis-probing research -- in particular once LHC finds New Physics directly -- while charm and $\tau$ studies are of the hypothesis-generating variety. 

Let me add one look at the `Grand Picture'. Heavy flavour studies continue to be of 
fundamental importance, its lessons cannot be obtained any other way, thus they cannot become 
obsolete and they can sweep out dynamical scales up to the 100 TeV domain, i.e. well beyond the direct reach of the LHC. The LHC is and has to be the centerpiece of our efforts for quite a while to come. Yet it has three natural daughters: the `straight daughter' or ILC; the `cinderella' or tau-charm factory; the `beautiful daughter' or Super-Flavour Factory 
$e^+e^- \to \Upsilon (4S,5S) \to b \bar b, c \bar c, \tau^+\tau^-$ with a luminosity of around 
$10^{36}$ cm$^{-2}$ s$^{-1}$. The latter will provide a data base of the required size and quality 
not only to make precise measurements, but more importantly to interpret them accurately.  We are at the beginning of a most exciting adventure, where we can be certain to find exciting new phenomena -- and we are most privileged to participate. 

\section*{Acknowledgments} 
Krakow is one of the truly great cities of the world not merely because of the beauty of its architecture, the civic sense of its citizens or its long history per se -- but also for a reason touching us more directly as citizens of academia: its Jagiellonian University founded in 1364 is practically a founding and certainly an elite member of the academy in central Europe. It has done our ideals and aspirations proud -- despite having had neighbours that all too often were less than benign. I consider it 
always a privilege 
to participate in a scientific meeting in Krakow, and I am grateful to the organizers -- my colleagues and friends -- to have granted me this privilege again. 
This work was supported by the NSF under grant PHY03-55098.

\appendix

\section{Calibrating and Validating Our Theoretical Tools}
\label{CALI}

 Theory actually faces two types of challenges: 

$\bullet$ 
{\em Generic} TeV scale New Physics scenarios would already have manifested themselves, in 
particular  through flavour changing neutral currents, since those are so highly suppressed 
within the SM. Apparently we are missing an important message about flavour dynamics -- this is 
the `New Flavour Problem'. The fact that studies of heavy flavour decays represent largely 
hypothesis-generating rather then  $\sim$-probing research is illustrated by the common use of classifications like `minimal-flavour-violation', `next-to-minimal-flavour-violation' etc. 

$\bullet$ 
To obtain precise SM predictions and likewise to interprete the data in a reliable way we have to bring 
nonperturbative QCD under theoretical control. This is the challenge I will address below. 

\subsection{~Heavy Quark Theory}
\label{HQTH}

Heavy quark theory based on heavy quark symmetry and heavy quark expansions (HQE) in 
{\em inverse} powers of the heavy quark mass $m_Q$ 
-- thus combining a global symmetry with a dynamical treatment -- is one 
of the most active and quickly progressing fields of QCD, although it is not often appreciated 
by the rest of the QCD community. Its central tool is the operator product expansion (OPE), which 
expresses (mostly inclusive) observables through a series of expectation values of 
local heavy flavour operators $O_i$ with coefficients that can be calculated in short distance dynamics: 
\beq 
{\rm observable}(H_Q \to f) = \sum_i c_i(f)\matel{H_Q}{O_i}{H_Q} 
\eeq 
The $c_i(f)$ contain in particular the CKM parameters and $m_Q$; more specifically 
the coefficients of the higher dimensional operators are suppressed by increasing powers 
of $1/m_Q$ (or the energy release $1/(m_b - m_c)$ for $b\to c$ transitions). The sometimes heard 
statement that the underlying concept of quark-hadron duality represents an {\em additional ad-hoc} 
assumption is not even wrong -- it just misses the point, as explained in considerable detail in 
Ref.\cite{VADE}. There it had been predicted that limitations to duality in evaluating $\Gamma_{SL}(B)$ 
cannot exceed $0.5 \%$.  

As far as {\em fully integrated} widths are concerned, the leading nonperturbative corrections 
are $\sim {\cal O}(1/m_Q^2)$ rather than ${\cal O}(1/m_Q)$ as is the case for hadronic masses 
and differential distributions \cite{BUV}. This result, which is intimately connected with colour being 
a locally gauged quantum number, is essential  for the goal of high accuracy: for $Q=b$ one then has 
nonperturbative corrections of order $(\Lambda_{NP}/m_b)^2 \sim (1/5)^2 = 0.04$, and one needs to 
control them merely on the 20\% level to achieve an overall accuracy of 1 \%.  Furthermore 
it is not necessarily foolish to apply a HQE to charm widths to obtain at least semiquantitative 
predictions. 

\subsection{~Applications to Semileptonic \& Radiative $B$ Decays}
\label{SLBDEC}

It has been demonstrated that high numerical accuracy can be achieved in our theoretical 
description \cite{BENSON1}. From inclusive semileptonic and radiative $B$ decays one has inferred \cite{BUCH}
\bea 
m_b^{\rm kin} (1\; \GeV) &=& (4.59 \pm 0.04) \GeV  \; \; \; \hat = \; \; \pm \; 1.0 \%\\
m_c^{\rm kin} (1\; \GeV) &=& (1.14 \pm 0.06) \GeV  \; \; \;  \hat =  \; \; \pm \; 5.0 \% \\
|V(cb)|_{\rm incl} &=&  (41.96 \pm 0.67) \cdot 10^{-3} \; \; \; \hat =  \; \; \pm \; 1.6 \%
\eea 
where the last line should be compared with what we know about the Cabibbo angle 
studied for a much longer period: 
\beq 
|V(us)| = 0.2252 \pm 0.0022  \; \; \; \hat = \; \;  \pm \; 1.0 \%
\eeq
The robust theoretical framework required for such achievements has been provided by 
Heavy Quark Theory \cite{HQT} implemented through the Wilsonian OPE and augmented 
by SV and other sum rules. Yet equally important was the impact 
of high quality data on total rates and distributions allowing to measure 
energy and hadronic mass moments in $B\to l \nu X_c, \, \gamma X_s$.  

To achieve such accuracy levels one had to determine and even define the 
heavy quark mass very carefully, since the weak decay widths depend on the fifth power of it. 
A few concise comments on this complex issue have to suffice here. 
(i) The {\em pole} mass cannot be used, since renormalon effects due to its infrared instability in full QCD induce irreducible uncertainties parametrically larger than the leading nonperturbative 
corrections. (ii) The $\overline{MS}$ mass is most appropriate, when the relevant scales are larger 
than $m_Q$ like in $Z\to b \bar b$ or $H \to b \bar b$; yet it is ill-suited for $H_Q$ decays, where the 
relevant scales are necessarily lower than $m_Q$. On the other hand it can be employed  
as a reference point for $m_Q$ extracted from different processes. 
(iii) The {\em kinetic} mass is defined by the scale dependence 
\beq 
dm_Q(\mu )/d\mu = - 16 \alpha _S(\mu)/3\pi + ... 
\eeq 
i.e., a linear scale dependence in the IR region. It is the most appropriate quantity for 
$H_Q$ decays, and its framework has been well developed now. 
(iv) The {\em 1 S} mass has principal short comings as explained in Ref.\cite{URI1S}. I view 
it as inferior to the {\em kinetic} mass. Therefore  
I was quite surprised -- to put it very mildly -- that PDG has declared by `ordre du mufti' 
to list only the {\em 1S}, but not the {\em kinetic} mass. I would be most grateful if somebody could explain 
to me, what the {\em scientific} reason behind this decision is.

Experimental cuts have often to be imposed on kinematical variables, like on the 
lepton and photon energies in $B \to l \nu X$ and $B \to \gamma X$, respectively. Those 
can create a 
serious theoretical problem though: for they reduce the amount of averaging over hadronic channels 
and thus might reduce the quantitative validity of quark-hadron duality; they manifestly introduce another energy scale potentially making the application 
of the OPE more ambiguous. This issue has been addressed theoretically concerning the lower cut 
$E_{\rm cut}$ on the photon energy in $B \to \gamma X$. Ignoring the sensitivity to  
$E_{\rm cut}$ will distort the spectrum, yet such `biases' can be corrected, and the validity of the OPE 
thus extended \cite{BENSONBIAS}. These predicted effects have been found in the data 
\cite{BUCH}. Analogous complications are expected to arise, when one measures lepton energy and hadronic mass moments in $B \to l \nu X_c$ with the lepton energy cut exceeding 1.6 GeV. 
{\em It would 
be most instructive to study, how the measured and predicted moments deviate from each other for 
$E_{\rm cut}^{\rm lept} \geq 1.6$ GeV.} The philosophy here is similar to that of engineers, who strain an engine to the breaking point to test its reliability. 

The lessons we are learning from  $B\to l \nu X_c, \gamma X$ help us with extracting 
$|V(ub)|$ from $B\to l \nu X_u$ in general as well as specific ways. Most importantly one does not 
need to `re-invent the wheel': it is one of the strengths of the OPE that the {\em same} HQP are 
to be used in both $B \to l \nu X_{c,u}, \, \gamma X$ and the nonleptonic rates, albeit with coefficients 
specific to the final state. $\Gamma (B \to l \nu X_u)$ can actually be calculated in terms 
of $|V(ub)|$ with higher accuracy than $\Gamma (B \to l \nu X_c)$. The problem arises on the 
experimental side, since the total width cannot be measured directly; severe cuts have to be 
imposed on kinematical variables to extract a signal from the huge $B\to l \nu X_c$ background. 
While the lepton energy endpoint region  provides a clean signal for $|V(ub)| \neq 0$, interpreting 
it with high numerical reliability is quite another matter. At least one has to perform such an analysis separately for $B_d$ and $B_u$ decays, since they are affected differently by weak annihilation \cite{WA}. Theoretically more promising ways are to analyze the hadronic recoil spectrum 
in $B \to l \nu X$ with and without cuts on $q^2$ \cite{DIKE,BAUER}. While I am skeptical about the accuracy 
presently claimed, I am optimistic that we can achieve {\em defensible} 5\% (or even better) precision 
over the next few years. 

\subsection{~On the Powers of the Dalitz Plot}
\label{DALITZ} 

Bringing hadronization under theoretical control obviously represents a stiff challenge. 
Yet since hadronization also enhances many signals for \cp~violation \cite{RIO}, 
we should view it as an essential even if 
quirky ally we can deal with by treating rich and complex data {\em judiciously}. Rather than viewing 
the Dalitz plot method  as a prehistoric remnant used by people too old to learn C++, it should be recognized as a mature and powerful high-sensitivity tool.  It will be crucial in saturating the discovery potential in $B$ (and $D$) decays, as sketched by a few examples. 

{\em Case I}: The angle $\phi_3$ in the CKM unitarity triangle is being extracted from 
$B^{\pm} \to D^{\rm neut} K^{\pm}$, where the neutral $D$ mesons have been identified  
through (a) flavour-specific or (b) ~-nonspecific modes, i.e. those common to $D^0$ and $\bar D^0$.  
Originally  one had considered only two-body channels for the latter. A new level of accuracy 
and reliability has been reached by relying on a full Dalitz plot analysis of 
$D^0/\bar D^0 \to K_S \pi^+\pi^-$ as pioneered by BELLE \cite{BART}. It requires a very substantial effort, yet this investment pays handsome profits in the long run, for the very complexity of a full Dalitz plot with its many correlations 
provides a profound quality check thus giving us confidence in the weak parameters extracted. Increases in statistics therefore translate into a largely commensurate gain in information 
with a {\em defensible} estimate of the uncertainties. 

{\em Case II}: In extracting the angle $\phi_2$ from \cp~asymmetries in $B\to \pi's$ one has to deal 
with the complication of two quark-level operators contributing, namely a tree as well as a 
(one-loop) Penguin one. 
The theoretically cleanest, yet experimentally very challenging method is based on 
analyzing $B_d \to \pi^+\pi^- /2\pi^0$ \& $B^{\pm}\to \pi^{\pm}\pi^0$. A recent favourite 
has been to study $B \to 2\rho$ \cite{BART}. Experimentally it offers some advantages, yet theoretically suffers 
from significant drawbacks as well. For those transitions have to be inferred from 
$B \to 4 \pi$, which will contain final states other than $2\rho$, namely $\rho \sigma$, 
$2\sigma$ etc., where it does not matter, whether the $\sigma$ is a bona fide resonance or not. 
One should note that even if the $\sigma$ is a genuine resonance, it can{\em not} be 
described by a simple Breit-Wigner excitation function \cite{ULF}. 
Imposing a cut on the di-pion mass is not overly selective due to the large $\rho$ width.  For as 
stated repeatedly above we have to aim for an accuracy level of very few percent at most. 
Similar concerns affect the analysis based on $B \to \rho \pi$, since the primary reaction 
$B \to 3\pi$ contains coherent contributions from $B \to \sigma \pi, [3\pi]_{NR}$ as well 
\cite{ULF}. Ultimately 
the goal has to be to perform full Dalitz plots analyses \cite{ART} of $B_d \to \pi^+\pi^- \pi^0/3\pi^0$ \& 
$B^{\pm} \to \pi^{\pm} \pi^+\pi^-/ \pi^{\pm}\pi^0\pi^0$, where the multineutral final states provide an 
important cross check. In doing so one has to implement all the constraints from chiral dynamics applied to $\pi\pi$ scattering. 

{\em Case III}: The angle $\phi_1$  can be extracted also from $B_d \to \phi K_S$. Within the SM this Penguin driven channel has to exhibit basically the same 
\cp~asymmetries as in $B_d \to \psi K_S$, where it has been determined with high precision. 
Any deviation signals the intervention of New Physics, which actually finds fertile ground in 
$B_d \to \phi K_S$: its SM amplitude is considerably suppressed; 
the transition is driven by a single operator; we have a reliable SM prediction and finally, the 
$\phi$ constitutes a narrow resonance. The BELLE/BABAR average yields a value for the 
$S$ term for $B_d \to \phi K_S$ and analogous modes like $B_d \to \eta K_S$ that is somewhat low 
compared to the SM prediction, yet not inconsistent with it \cite{ICHEP06}. I am most intrigued and tantalized 
by it, since the present central values are in a most natural range for New Physics. Again ultimately 
one has to analyze time-{\em dependent} Dalitz plots for $B_d \to K^+K^-K_S$ (and cross reference 
them with $B_d \to 3K_S$ as well as $B^{\pm} \to K^{\pm}K^+K^-/K^{\pm}K_SK_S$). One should also 
note that within the SM $B_d \to f(980)K_S$ has to exhibit a \cp~asymmetry of equal magnitude, yet 
opposite sign to $B_d \to \phi K_S$, since the two final states have opposite \cp~parity. That means 
that a $B_d \to f(980)K_S$ {\em amplitude} 10 \% the size of that for $B_d \to \phi K_S$ -- thus quite 
insignificant in rate -- would reduce the observable \cp~asymmetry by 20 \%. 

\subsection{~$B\to \tau \nu D$ vs. $B\to \mu \nu D$}
\label{BPSAP}

While 
$\Gamma (B \to e \nu D)$ is dominated by a single form factor $f_+$,  
$\Gamma (B \to \tau \nu D)$  is affected also by the second form factor $f_-$, since $m_{\tau}$ 
is not negligible on the scale of $M_B - M_D$; secondly, the range of $q^2$, which forms the 
argument in $f_{+,-}(q^2)$ is quite different for the two transitions. This complication can however be 
overcome by a novel theoretical tool, namely Uraltsev's `BPS' approximation. Applying the latter 
to $B \to e \nu D$ should allow to extract  $|V(cb)|$ with very few percent uncertainty. 
Once this approximation has been validated by comparing $|V(cb)|_{BPS}$ with 
$|V(cb)|_{incl}$, it can be relied upon for calculating the SM value for 
$\Gamma (B\to \tau \nu D)/\Gamma (B \to e \nu D)$.

\end{document}